\documentclass[onecolumn,pra,notitlepage]{revtex4-1}
\usepackage{amsmath}
\usepackage{graphicx}
\usepackage{wrapfig}
\usepackage{enumerate}
\usepackage{float}
\usepackage{standalone}
\usepackage{lineno}
\usepackage{subcaption}

\begin{document}

\title{Cavity optomechanics with a suspended resonant mirror}

\author{Trishala Mitra$^1$, Gurpreet Singh$^1$, Søren Peder Madsen$^2$ and Aur\'{e}lien Dantan$^1$}\email{dantan@phys.au.dk} 

\address{$^1$Department of Physics and Astronomy, Aarhus University, DK-8000 Aarhus C, Denmark\\$^2$Department of Mechanical and Production Engineering, Aarhus University, DK-8000 Aarhus C, Denmark}
\begin{abstract}
We investigate optomechanical effects in cavities consisting of a plane-plane arrangement of a broadband reflectivity mirror and an ultrathin, suspended resonant mirror possessing a high-Q internal optical resonance. We first investigate dispersive and photothermal optomechanics in such cavities on the basis of a generic analytical model as well as finite element method simulations of realistic structures. We then report on experimental optical spring measurements using a suspended silicon nitride membrane patterned with a subwavelength grating. The observed optical spring variations match those expected from the theoretical model, but with a magnitude several hundreds times larger than expected from the dispersive optomechanical coupling only, thus suggesting strong photothermal optomechanical coupling.
\end{abstract}

\date{\today}

\maketitle

%%%%%%%%%%%%%%%%%%%%%%%%%%  body  %%%%%%%%%%%%%%%%%%%%%
\section{Introduction}

The coupling via radiation pressure of the motion of flexible ultrathin membranes in optical resonators has enabled fundamental advances in the field of cavity optomechanics~\cite{Aspelmeyer2014}. The use of low-reflectivity membranes in high-finesse optical resonators in the membrane-in-the-middle geometry~\cite{Thompson2008,Wilson2009} has led to the observation of the radiation pressure shot-noise~\cite{Purdy2013}, the generation of ponderomotive squeezing~\cite{Purdy2013b}, the observation of optomechanically induced transparency~\cite{Karuza2013}, topological energy transfer~\cite{Xu2016}, motional ground state cooling~\cite{Saarinen2023}, the control of quantum motion by measurement~\cite{Rossi2018}, and optomechanical heat transport and long-range interactions between resonators~\cite{Yang2020b,Yao2025}, among others. 

Recently, there has also been substantial progress in increasing the otherwise relatively low reflectivity of thin micromechanical membranes by patterning them with one- and two-dimensional photonic crystal structures~\cite{Kemiktarak2012,Bui2012,Kemiktarak2012a,Norte2016,Reinhardt2016,Chen2017,Nair2019}, ultimately allowing them to be applied as end-mirrors in optomechanical cavities~\cite{Sang2022,Xu2022,Zhou2023,Enzian2023,Khokhar2026}.

Since the increase in reflectivity at specific wavelengths of nanopatterned thin films exploits the interference between the incident light and guided modes in the structured film, it is also possible to design---and take advantage of---potentially narrow optical internal (\textit{Fano}) resonances of the resonant mirror and thereby realize {\it Fano} cavities with strongly wavelength-dependent reflectors~\cite{Naesby2018,Cernotik2019,Mitra2024,Kirkegaard2025}. Such narrow-linewidth, ultracompact microcavities may open new possibilities for cavity optomechanics~\cite{Cernotik2019,Fitzgerald2021,Peralle2024,Du2026} and, recently, strong and unconventional optomechanical interactions have been experimentally observed with a photonic crystal semiconductor membrane suspended over a Bragg reflector~\cite{Manjeshwar2023}. Both dispersive and photothermal optomechanical interactions have also been observed with suspended photonic crystal membranes~\cite{Woolf2013}.

In this work we investigate cavity optomechanics for resonators consisting of a plane-plane arrangement of a broadband reflectivity mirror and a suspended, resonant mirror. We first investigate Fano cavity optomechanics in this canonical system on the basis of the analytical coupled-mode model of Ref.~\cite{Cernotik2019} and including both dispersive and photothermal effects. We show in particular that, in the unresolved sideband regime, the magnitude of the dispersive optical spring shift in a Fano cavity is the same as in the corresponding broadband mirror cavity with the same internal loss, in spite of the linewidth reduction of the Fano cavity observed at short cavity lengths. We then corroborate these generic findings with finite-element simulations of realistic structures, such as those used in the experimental investigations. Experimentally, we make use of a highly reflective silicon nitride membranes possessing a high-Q optical resonance to realize Fano cavities with lengths ranging from a few to a few hundreds of microns, and measure the optical spring shift as a function of the input power and cavity detuning. The observed optical spring variations match those expected from the Fano cavity optomechanics model, but with a magnitude several hundred times larger than expected from the dispersive optomechanical coupling only, thus suggesting strong photothermal optomechanical coupling.

%%%%%%%%%%%%%%%%%%%%%%%%%%%%%%%%%%%%%%%%%%%%%%%%%%%%

\section{Coupled-mode dispersive and photothermal optomechanics model}
\label{sec:theory}

\subsection{Broadband mirror cavity}

\begin{figure}[h!]
\centering\includegraphics[width=\textwidth]{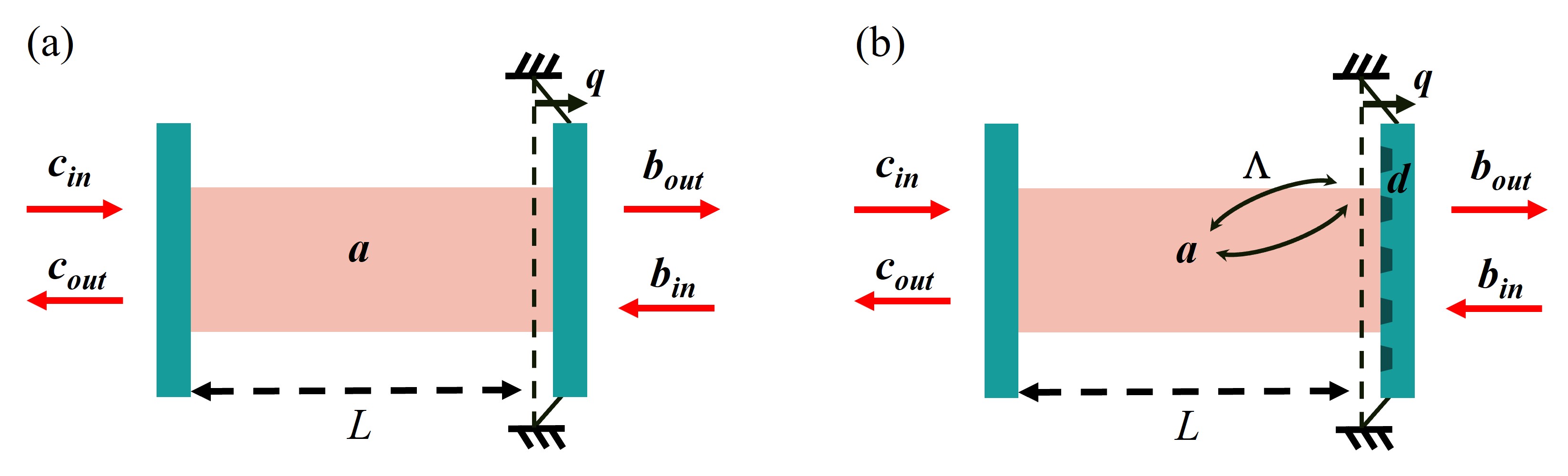}
\caption{Schematics of broadband mirror (a) and Fano mirror (b) optomechanical cavities. In both cavities the intracavity mode $a$ couples dispersively to the movable mirror via radiation pressure. In the Fano mirror cavity the intracavity mode is in addition coupled to the Fano mirror mode $d$.}
\label{fig:schematic}
\end{figure}

\subsubsection{Equations of motion}

We start by considering a linear Fabry-Perot cavity with two broadband mirrors, as depicted in Fig.~\ref{fig:schematic}(a). In absence of light, the cavity length is $L$ and the right-hand mirror can oscillate around its equilibrium position with a frequency $\omega_m$. We assume that, when light with frequency $\omega_L$ is injected into the cavity, its motion is coupled to the intracavity field via the standard dispersive linear optomechanical Hamiltonian
\begin{equation}
H_\textrm{disp}=-\hbar Ga^\dagger a q,
\end{equation} where $a$ and $a^\dagger$ are the annihilation and creation operators for the intracavity field mode with frequency $\omega_0$, $q$ the mechanical displacement operator and
\begin{equation}
G=\frac{\omega_0}{L}q_0
\end{equation}
is the single-photon optomechanical coupling, where $q_0=\sqrt{\hbar/2m\omega_m}$ is the zero-point motion amplitude for the mechanical mode with mass $m$.

To introduce photothermal effects we supplement this purely dispersive linear optomechanical coupling with a standard phenomenological model of photothermal optomechanics~\cite{Metzger2004,Pinard2008,Restrepo2011,Shakespeare2024}. We thus consider a total optomechanical force which is the sum of the radiation pressure force $F_\textrm{disp}=\hbar Ga^\dagger a$ and a photothermal force $F_\textrm{ph}$ proportional to the intensity absorbed by the structure and with a time delay $\tau$,
\begin{align}
F_\textrm{tot}&=F_\textrm{disp}+F_\textrm{ph}=\hbar Ga^\dagger(t) a(t)+\hbar \beta G\int_{-\infty}^t \frac{e^{-(t-t')/\tau}}{\tau}a^\dagger(t')a(t')dt',
\end{align}
where $\beta$ represents the relative magnitude of the photothermal force with respect to the radiation pressure force (if $a^\dagger a$ was constant in time, $F_\textrm{ph}=\beta F_\textrm{disp}$). $\beta$ encompasses photon absorption/loss in the structure as well as its transduction into mechanical motion, and can be positive or negative, depending on the optomechanical structure considered~\cite{Metzger2004,Metzger2008,Metzger2008b,Pinard2008,Restrepo2011,Woolf2013,Shakespeare2024}. $\tau$ represents the characteristic time of the photothermal dynamics due to e.g. thermal relaxation. Both parameters are in general difficult to extract from first principles for a given structure.

The Heisenberg-Langevin equations of motion describing the dynamics of the cavity field and the mechanical mode can be written as~\cite{Cernotik2019}
\begin{align}
\dot{a}&=-(\kappa_R+\kappa_0+\kappa_0'+i\Delta)a+iGaq+\sqrt{2\kappa_R}c_\textrm{in}+\sqrt{2\kappa_0}b_\textrm{in}+\sqrt{2\kappa_0'}b'_\textrm{in},\\
\dot{p}&=-\gamma_mp-\omega_mq+F_m+Ga^\dagger a+\beta G\int_{-\infty}^t \frac{e^{-(t-t')/\tau}}{\tau}a^\dagger(t')a(t')dt',\label{eq:p_disp}\\
\dot{q}&=\omega_m p,
\end{align}
where $\kappa_R=\kappa_L T_R/4$ and $\kappa_0=\kappa_L T_0/4$ are the field decay rates through the input and output mirrors with transmission losses $T_R$ and $T_0$, respectively, $\kappa_0'=\kappa_L A/4$ the decay rate associated with the single roundtrip absorption losses $A$, $\kappa_L=c/L$ is twice the cavity free spectral range (FSR), $\Delta=\omega_0-\omega_L$ is the cavity detuning, $c_\textrm{in}$ and $b_\textrm{in}$ represent the input field modes from the left- and right-hand side of the cavity, $b'_\textrm{in}$ the input mode associated with the absorption losses, $p$ is the mechanical momentum ($[q,p]=i\hbar$), $\gamma_m$ is the mechanical damping rate and $F_m$ the associated thermal noise force. The finesse of the cavity is then $2\pi/(T_R+T_0+A)$.

\subsubsection{Steady state and fluctuations}

We then proceed in a standard fashion by linearizing the fluctuations of the operators around their classical steady state mean values (e.g. $a=\bar{a}+\delta a$). Assuming that light is injected from the left-hand side only ($\bar{b}_\textrm{in}=0$, $\bar{b}'_\textrm{in}=0$) with a mean photon flux $|\bar{c}_\textrm{in}|^2=P_\textrm{in}/\hbar\omega_0$, where $P_\textrm{in}$ is the input power, and neglecting the small, static radiation-pressure induced displacement, the mean intracavity field amplitude is given by
\begin{equation}
\bar{a}=\frac{\sqrt{2\kappa_R}}{\kappa_B+i\Delta}\bar{c}_\textrm{in},
\end{equation}
where $\kappa_B=\kappa_R+\kappa_0+\kappa'_0$ is the total intracavity field decay rate and the steady state cavity transmission $\mathcal{T}_\textrm{cav}=|\bar{b}_\textrm{out}/\bar{c}_\textrm{in}|^2$ can be readily shown to be given by 
\begin{equation}
\mathcal{T}_\textrm{cav}=\frac{4\kappa_R\kappa_0}{\kappa_B^2+\Delta^2}
\label{eq:TcavB}
\end{equation}

The Fourier transform of the equation of motion for the intracavity field fluctuations reads
\begin{equation}
\epsilon_a(\omega)^{-1}\delta a(\omega)=iG\bar{a}\delta q(\omega)+\sqrt{2\kappa_R}\delta c_\textrm{in}(\omega)+\sqrt{2\kappa_0}\delta b_\textrm{in}(\omega)+\sqrt{2\kappa'_0}\delta b'_\textrm{in}(\omega),
\end{equation}
where $\delta o(\omega)$ denotes the Fourier transforms of $\delta o$ and the $a$ mode susceptibility
\begin{equation}
\epsilon_a(\omega)^{-1}=\kappa_B+i(\Delta-\omega)
\end{equation}
has been introduced. Without loss of generality one can assume further $\bar{a}$ to be real (by redefining the input field phase). The fluctuations of the radiation force are then given by (see e.g.~\cite{Metzger2004,Pinard2008,Restrepo2011,Shakespeare2024})
\begin{align}
\delta F_\textrm{tot}(\omega)=\hbar G\bar{a}\left(1+\frac{\beta}{1-i\omega\tau}\right)(\delta a(\omega)+\delta a^\dagger(\omega)).
\end{align}
Using $\delta a(\omega)=iG\bar{a}\epsilon_a(\omega)\delta q(\omega)+...$ and $\delta a^\dagger(\omega)=-iG\bar{a}\epsilon_a^*(-\omega)\delta q(\omega)+...$, where "..." represent terms proportional with the input noise modes (which we omit for the sake of clarity), leads to an equation of motion for the Fourier transform of the mechanical displacement
\begin{equation}
\label{eq:deltaq}
\left[\omega_m^2-\omega^2-i\gamma_m\omega\right]\delta q(\omega)=i\omega_mG^2\bar{a}^2\left(1+\frac{\beta}{1-i\omega\tau}\right)\left(\epsilon_a(\omega)-\epsilon_a^*(-\omega)\right)\delta q(\omega)+...,
\end{equation}
In the weak optomechanical coupling /good mechanical oscillator regimes relevant for this work, this amounts to an effective damped harmonic oscillator response for the mechanics
\begin{equation}
\label{eq:mech}
\left[\omega_m'^2-\omega^2-i\omega\gamma_m'\right]\delta q(\omega)=...
\end{equation}
where $\omega_m'=\omega_m+\delta\omega_m$ and $\gamma_m'=\gamma_m+\delta \gamma_m$ are the effective mechanical frequency and damping given by
\begin{align}
\delta\omega_m&=\tfrac{1}{2}G^2\bar{a}^2\textrm{Im}\left[\left(1+\frac{\beta}{1-i\omega_m\tau}\right)\left(\epsilon_a(\omega_m)-\epsilon_a^*(-\omega_m)\right)\right],\label{eq:deltaomegamB}\\
\delta\gamma_m&=G^2\bar{a}^2\textrm{Re}\left[\left(1+\frac{\beta}{1-i\omega_m\tau}\right)\left(\epsilon_a(\omega_m)-\epsilon_a^*(-\omega_m)\right)\right].\label{eq:deltagammamB}
\end{align}
While these are both proportional to the intracavity power, their variations with the detuning can be quite complex depending on the magnitude and sign of $\beta$, on the product $\omega_m\tau$, as well as on the mechanical frequency to cavity linewidth ratio $\omega_m/\kappa_B$. In the following we focus on the unrevolved sideband ($\kappa_B\gg\omega_m$) and fast thermalization $(\omega_m\tau\ll 1$) regimes which are relevant for the experiments presented in Sec.~\ref{sec:experiment}.

\subsubsection{Unresolved sideband regime}

In absence of photothermal effects ($\beta=0$) and in the unresolved sideband regime ($\kappa_B\gg\omega_m$), one retrieves the well-known expression for the mechanical resonance frequency (or \textit{optical spring}) shift 
\begin{equation}
\delta\omega_m\simeq -G^2\bar{a}^2\frac{\Delta}{\kappa_B^2+\Delta^2}=-2G^2|\bar{c}_\textrm{in}|^2\kappa_R\frac{\Delta}{(\kappa_B^2+\Delta^2)^2}.
\label{eq:springB}
\end{equation}
The amplitude of the optical spring varies linearly with the input power and dispersively with the detuning, and its magnitude is largest when $|\Delta|=\kappa_B/\sqrt{3}$. Interestingly, the maximum optical spring amplitude,
\begin{equation}
\delta\omega_m^{\textrm{(max)}}=G^2|\bar{c}_\textrm{in}|^2\frac{3\sqrt{3}}{8}\frac{\kappa_R}{\kappa_B^3}=\eta^2|\bar{c}_\textrm{in}|^26\sqrt{3}\frac{T_R}{(T_R+T_0+A)^3},
\label{eq:deltaomegamaxB}
\end{equation}
where $\eta=(2\pi/\lambda_0) q_0$ is the Lamb-Dicke parameter, is independent of the cavity length.

\begin{figure}[h]
\centering
\includegraphics[width=\columnwidth]{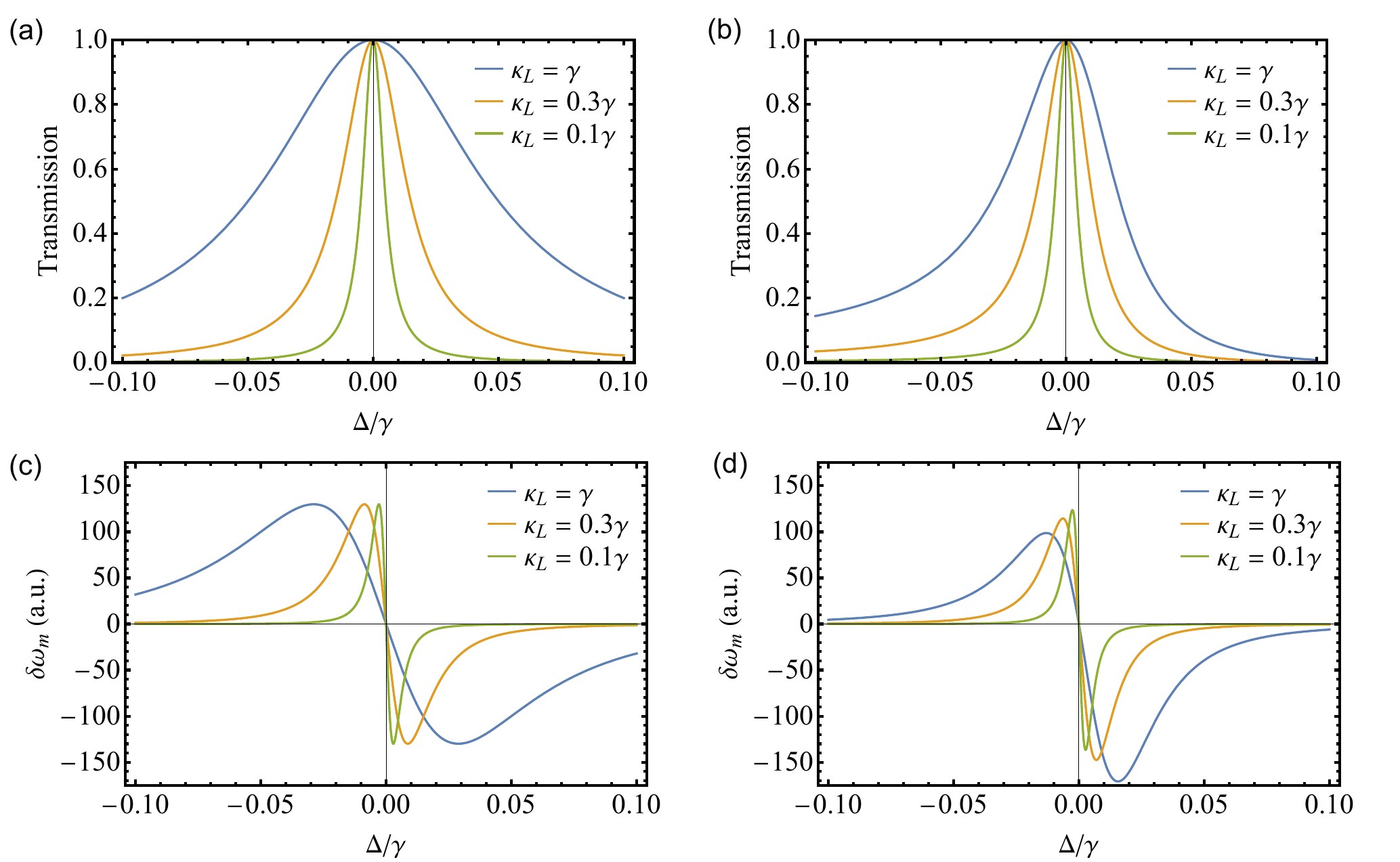}
\caption{Cavity transmission spectra of a broadband mirror (a) and Fano mirror cavity (b) for different cavity lengths, corresponding to different $\kappa_L/\gamma$ ratios ($\kappa_L=\gamma$: blue, $\kappa_L=0.3\gamma$: orange, $\kappa_L=0.1\gamma$: green). The other parameters are $T_R=T_0=0.1$, $A=0$. (c) and (d): corresponding optical spring shifts ($\eta$ and $|\bar{c}_\textrm{in}|^2$ are taken to be unity in both plots to facilitate the relative comparison).}
\label{fig:fig1}
\end{figure}

An illustration of the analytical predictions of Eqs.~(\ref{eq:TcavB}) and (\ref{eq:springB}) is given in Figs.~\ref{fig:fig1}(a) and (c), which show the broadband mirror cavity transmission spectrum and the corresponding optical spring shift as a function of the detuning for different cavity lengths.

When photothermal effects are present ($\beta\neq 0$) and in the fast thermalization regime ($\omega_m\tau\ll 1$), the optical spring variations are similar, i.e. given by Eq.~(\ref{eq:springB}), albeit multiplied by a factor $\xi\simeq 1+\beta$.

\subsection{Fano cavity}

\subsubsection{Equations of motion}

To investigate dispersive and photothermal optomechanics in a Fano cavity consisting of a broadband mirror and a resonant mirror, as depicted in Fig.~\ref{fig:schematic}(b), we make use of the coupled-mode model of Ref.~\cite{Cernotik2019} and introduce a Fano mirror mode $d$, whose coupling with the $a$ mode and the input/output modes describes the interference leading to the internal optical resonance~\cite{Fan2003}. Following the notations of Ref.~\cite{Cernotik2019} and under the same linear optomechanical coupling assumptions as previously, the equations of motion for the optical and mechanical modes now read
\begin{align}
\dot{a}=&-(\kappa_L+\kappa_R+\kappa'_0+i\Delta_a)a-\mathcal{G}d+iGaq+\sqrt{2\kappa_L}b_\textrm{in}+\sqrt{2\kappa_R}c_\textrm{in}+\sqrt{2\kappa'_0}b'_\textrm{in}, \\
\dot{d}=&-(\gamma+i\Delta_d)d-\mathcal{G}a+\sqrt{2\gamma}b_\textrm{in},\\
\dot{p}=&-\gamma_mp-\omega_m q+F_m+Ga^\dagger a+\beta G\int_{-\infty}^t \frac{e^{-(t-t')/\tau}}{\tau}a^\dagger(t')a(t')dt',\\
\dot{q}=&\quad\omega_m p,
\end{align}
with \begin{align}
\mathcal{G}&=\sqrt{\kappa_L\gamma}+i\sqrt{\kappa_0\gamma},\\
\Delta_a&=\Delta_d+2\sqrt{\kappa_0\kappa_L},
\end{align}
where $\gamma$ is the halfwidth of the Fano mirror resonance. As shown in Ref.~\cite{Cernotik2019}, the input and output modes are related to the intracavity and Fano modes by
\begin{align}
b_\textrm{out}&=b_\textrm{in}-\sqrt{2\kappa_L}a-\sqrt{2\gamma}d,\\
c_\textrm{out}&=c_\textrm{in}-\sqrt{2\kappa_R}a.
\end{align}

\subsubsection{Steady state and fluctuations}

Assuming as previously that the cavity is driven through the broadband mirror with a mean photon flux $|\bar{c}_\textrm{in}|^2=P_\textrm{in}/\hbar\omega_0$, and neglecting the backaction of the mechanics on the $a$ and $d$ modes, the steady state amplitudes of both modes are 
\begin{align}
\bar{d}&=-\frac{\mathcal{G}}{\gamma+i\Delta_d}\bar{a}=-\mathcal{G}\epsilon_d(0)\bar{a},\\
\bar{a}&=\frac{\sqrt{2\kappa_R}}{\epsilon_a(0)^{-1}-\mathcal{G}^2\epsilon_d(0)}\bar{c}_\textrm{in}=\sqrt{2\kappa_R}\tilde{\epsilon}_a(0) \bar{c}_\textrm{in},
\end{align}
where the (bare) susceptibilities of both modes have been introduced
\begin{align}
\epsilon_a^{-1}(\omega)&=\kappa_L+\kappa_R+\kappa'_0+i\Delta_a-i\omega,\\
\epsilon_d^{-1}(\omega)&=\gamma+i\Delta_d-i\omega.
\end{align}
and $\tilde{\epsilon}_a$ is the dressed susceptibility of the $a$ mode
\begin{align}
\tilde{\epsilon}_a&=\frac{1}{\epsilon_a^{-1}-\mathcal{G}^2\epsilon_d}.
\end{align}

The steady state transmission of the cavity $ \mathcal{T}_\textrm{cav}=|\bar{b}_\textrm{out}/\bar{c}_\textrm{in}|^2$ can be readily found to be
\begin{equation}
\label{eq:Tcav}
\mathcal{T}_\textrm{cav}=\left|\frac{\sqrt{2\kappa_R}[\epsilon_d(0)\mathcal{G}\sqrt{2\gamma}-\sqrt{2\kappa_L}]}{\epsilon_a^{-1}(0)-\mathcal{G}^2\epsilon_d(0)}\right|^2.
\end{equation}

The linearized equations of motion for the fluctuations of the $a$ and $d$ modes are
\begin{align}
\delta \dot{a}=&-(\kappa_L+\kappa_R+\kappa'_0+i\Delta_a')\delta a-\mathcal{G}\delta d+G\bar{a}\delta q+\sqrt{2\kappa_L}\delta b_\textrm{in}+\sqrt{2\kappa_R}\delta c_\textrm{in}+\sqrt{2\kappa'_0}\delta b'_\textrm{in},\label{eq:deltaa}\\
\delta \dot{d}=&-(\gamma+i\Delta_d')\delta d-\mathcal{G}\delta a+\sqrt{2\gamma}\delta b_\textrm{in},\label{eq:deltad}
\end{align}
where $\Delta_a'$ and $\Delta_d'$ are the detunings including the static mechanical displacement shift, which is typically small and which we will henceforth neglect. In Fourier space, Eqs.~(\ref{eq:deltaa}) and (\ref{eq:deltad}) lead to $\delta a(\omega)=iG\bar{a}\tilde{\epsilon}_a(\omega)\delta q(\omega)+...$ and $\delta a^\dagger(\omega)=-iG\bar{a}\tilde{\epsilon}^*_a(-\omega)\delta q(\omega)+...$, which subsequently lead to an expression similar to Eq.~(\ref{eq:mech}) for the effective mechanical response. The optomechanical frequency shift and (anti)damping are, respectively,
\begin{align}
\delta\omega_m=\tfrac{1}{2}G^2\bar{a}^2\textrm{Im}\left[\left(1+\frac{\beta}{1-i\omega_m\tau}\right)\left(\tilde{\epsilon}_a(\omega_m)-\tilde{\epsilon}_a^*(-\omega_m)\right)\right],\\
\delta\gamma_m=G^2\bar{a}^2\textrm{Re}\left[\left(1+\frac{\beta}{1-i\omega_m\tau}\right)\left(\tilde{\epsilon}_a(\omega_m)-\tilde{\epsilon}_a^*(-\omega_m)\right)\right],
\end{align}
i.e. formally identical to Eqs.~(\ref{eq:deltaomegamB})-(\ref{eq:deltagammamB}) with $\epsilon_a$ replaced by $\tilde{\epsilon}_a$.

\subsubsection{Unresolved sideband regime}

In the unresolved sideband and fast thermalization regimes, the previous expressions substantially simplify by evaluating the susceptibilities at zero frequencies, yielding a mechanical resonance frequency shift given by
\begin{align}
\label{eq:shift_exact}
\delta\omega_m= 2\kappa_R|\bar{c}_\textrm{in}|^2\xi G^2|\tilde{\epsilon}_a|^2\textrm{Im}[\tilde{\epsilon}_a],
\end{align}
where $\xi=1+\beta$. In the vicinity of the narrow Fano cavity resonance at $\Delta_d=0$, one has
\begin{align}
\tilde{\epsilon}_a&\simeq \frac{\gamma}{\kappa_L+\gamma}\frac{1}{\kappa_F'+i\Delta_d}
\end{align}
with
\begin{align}
\label{eq:kappaF}
\kappa_F'&=\kappa_F(1-\nu\Delta_d),\\
\nonumber\kappa_F&=\frac{\gamma(\kappa_R+\kappa_0+\kappa'_0)}{\kappa_L+\gamma}=\frac{\gamma\kappa_L}{\kappa_L+\gamma}\frac{T_R+T_0+A}{4},\\
\nonumber \nu &=\frac{2\sqrt{\kappa_0\kappa_L}}{\gamma(\kappa_R+\kappa_0+\kappa'_0)}=\frac{4}{\gamma}\frac{\sqrt{T_0}}{T_R+T_0+A},
\end{align}
resulting in a transmission function well-approximated by
\begin{align}
\label{eq:trans_ana}
\mathcal{T}_\textrm{cav}\simeq \frac{4T_0T_R}{(T_0+T_R+A)^2}\frac{1}{1+(\Delta_d/\kappa_F')^2}.
\end{align}
As illustrated in Fig.~\ref{fig:fig1}(b), the Fano mirror cavity transmission spectrum shows an asymmetric Fano profile with a HWHM given by $\kappa_F$, which coincides with the standard broadband cavity HWHM at long lengths, but is comparatively reduced at short lengths.

The optical spring shift is then given by
\begin{align}
\label{eq:shift_a_ana}\delta\omega_m&\simeq -\tfrac{1}{2}\xi G^2|\bar{c}_\textrm{in}|^2\kappa_LT_R\left(\frac{\gamma}{\kappa_L+\gamma}\right)^3\frac{\Delta_d}{(\kappa_F'^2+\Delta_d^2)^2}.
\end{align}
As shown in Fig.~\ref{fig:fig1}, the optical spring displays a slightly asymmetric dispersive profile, reflecting the asymmetric Fano cavity profile.

If one ignores the asymmetry of the cavity response around resonance, though, and approximates $\kappa_F'$ with $\kappa_F$ (Lorentzian cavity spectrum), the magnitude of the optical spring shift is found to be maximum when $|\Delta_d|\simeq\kappa_F/\sqrt{3}$ and the maximum optical spring shift then reduces to
\begin{align}
\delta\omega_m^\textrm{max}&\simeq \xi G^2|\bar{c}_\textrm{in}|^2T_R\kappa_L\left(\frac{\gamma}{\kappa_L+\gamma}\right)^3\frac{3\sqrt{3}}{32\kappa_F^3}=\xi\eta^2|\bar{c}_\textrm{in}|^26\sqrt{3}\frac{T_R}{(T_R+T_0+A)^3},
\end{align}
which coincides with Eq.~(\ref{eq:deltaomegamaxB}) obtained in the broadband cavity case. 

\subsection{Physical interpretation}

The cavity linewidth narrowing observed in a Fano cavity as compared to a broadband cavity with the same length and total losses can be interpreted as an increased effective length due to the interference of the intracavity photons with the guided mode of the Fano mirror. Indeed, neglecting the resonance asymmetry factor $\nu$ in Eq.~(\ref{eq:kappaF}), the Fano cavity linewidth can be recast as
\begin{equation}
\kappa_F=\frac{c\mathcal{L}/4}{L+L_0},
\end{equation}
where $\mathcal{L}=T_R+T_0+A$ is the total intracavity round-trip loss and $L_0=c/\gamma$ is the effective cavity length in the short cavity/ Fano regime. In that sense, $L_0$ also represents the characteristic cavity length over which the transition from the long to the short cavity regimes occurs.

The effective optomechanical coupling can be evaluated by estimating the shift in the cavity resonance frequency per displacement. If the Fano mirror is displaced from its equilibrium position by an amount $\bar{q}$, the steady state cavity transmission is readily shown to be given by Eq.~(\ref{eq:Tcav}) with $\epsilon_a(0)$ replaced by $\epsilon_a(G\bar{q})$. In the vicinity of the Fano resonance, the cavity transmission is still given by Eq.~(\ref{eq:trans_ana}), but with a shifted detuning
\begin{equation}
\Delta_d'=\Delta_d-G\frac{\gamma}{\kappa_L+\gamma}\bar{q}.
\end{equation}
The cavity resonance shift per displacement is then
\begin{equation}
-G\frac{\gamma}{\kappa_L+\gamma}=-\frac{\eta c}{L+L_0},
\end{equation}
which reduces to the standard broadband cavity dispersive optomechanical coupling in the long cavity regime ($L\gg L_0$), but becomes length-independent in the short cavity/Fano regime. This shows that the ratio of the dispersive OM coupling to the cavity linewidth is also length-independent in a Fano cavity. This is also consistent with the observation that the photons which predominantly populate the guided mode in the short cavity regime and which give rise to the linewidth narrowing do not contribute with a net longitudinal momentum exchange with the Fano mirror.

%%%%%%%%%%%%%%%%%%%%%%%%%%%%%%%%%%%%%%%%%%%%%%%%%%%%

\section{Numerical simulations}
\label{sec:simulations}

\subsection{Simulated Fano cavity}

\begin{figure}[h]
\centering
\includegraphics[width=0.32\columnwidth]{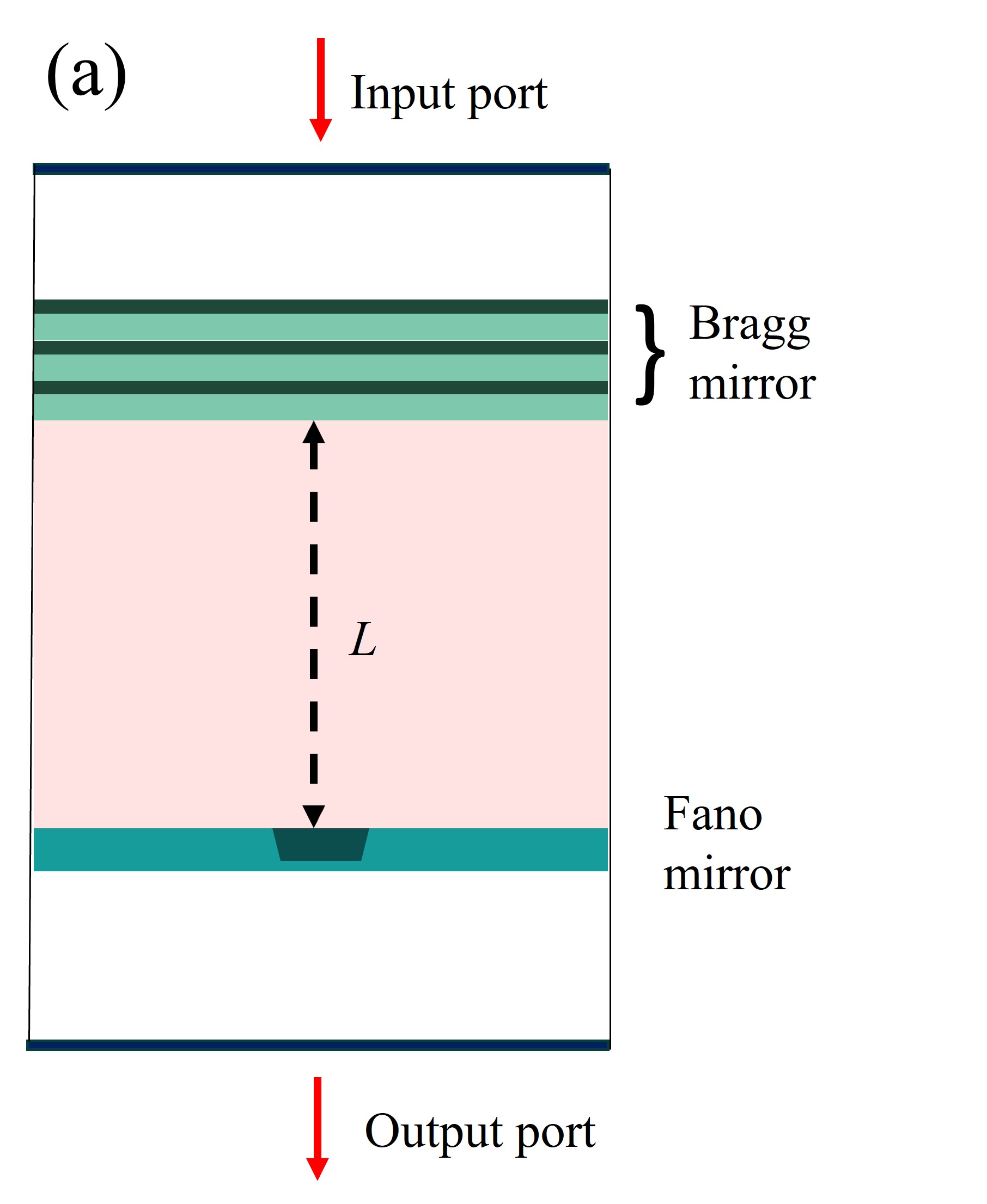}
\includegraphics[width=0.54\columnwidth]{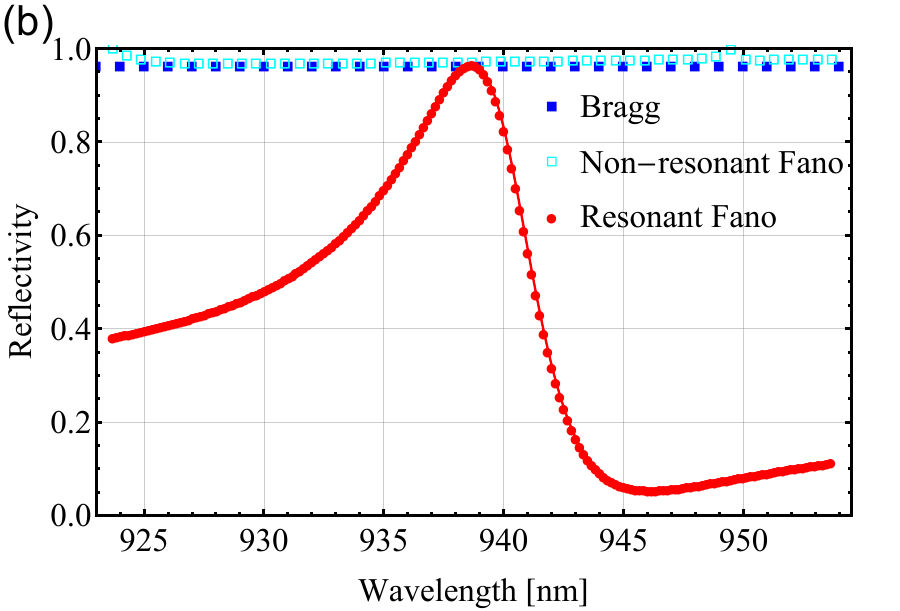}
\caption{(a) Plane wave model with horizontal periodic boundary conditions. (b) Simulated reflectivity spectra of the Bragg mirror (filled blue squares), the non-resonant Fano mirror (empty cyan squares) and the resonant Fano mirror (filled red dots). The red line shows the results of a fit with the reflectivity of a generic lossy Fano mirror~\cite{Mitra2024}, yielding a HWHM of 3 nm for the high-reflectivity resonance.}
\label{fig:PBCmodel}
\end{figure}

To test the accuracy of the predictions of the generic coupled-mode model with numerical simulations of a realistic structure similar to that investigated experimentally, we consider a cavity made of a broadband Bragg mirror and a thin subwavelength grating and simulate its transmission using Finite Element Method with COMSOL Multiphysics. 

The subwavelength grating consists of a one-dimensional 848 nm-periodic structure with quasi-rectangular grating fingers with width 508 nm and height 114 nm on an 89 nm-thick underlying layer. The grating fingers and underlying layer have a refractive index of 2.36. When illuminated at normal incidence by TM-polarized light, such a subwavelength grating displays a high-reflectivity Fano resonance at $\lambda_0=938.7$ nm with a half-width at half-maximum (HWHM) of 3 nm (Fig.~\ref{fig:PBCmodel}(b)). Such a structure possesses geometrical and optical properties similar to those of the "low-Q" Fano mirror used for the Fano cavity investigations of Ref.~\cite{Mitra2024} and allows for tractable FEM simulations when realistic grating and beam sizes are considered as in the Appendix.

The Bragg mirror consists of three pairs of lossless dielectric media with refractive index $n_1=1.5$ and $n_2=3.22$ and thicknesses $\lambda_0/4n_1$ and $\lambda_0/4n_2$, respectively, and displays a broadband reflectivity of 96\% over the range 920-960 nm (Fig.~\ref{fig:PBCmodel}(b)).

We then simulate the transmission of such Fano cavities for different cavity lengths chosen such that the cavity resonates at $\lambda_0$. We model cavities with infinite transverse extension illuminated by plane waves in Sec.~\ref{sec:PBCmodel} while the case of finite extension mirrors illuminated by Gaussian beams is tackled in the Appendix.

To simulate the transmission of the corresponding broadband cavity the refractive index of the subwavelength grating is artificially increased to 12.9, which results in a broadband high reflectivity of 96\% under normal-incidence, TM-polarized light illumination in the wavelength range considered (Fig.~\ref{fig:PBCmodel}(b)).

\subsection{PBC model}
\label{sec:PBCmodel}

The optomechanical coupling simulations are first carried out assuming periodic boundary conditions for the structure and plane-wave illumination. The dispersive optomechanical coupling is standardly evaluated by the evaluating the (linear) shift of the cavity resonance with the cavity length. Note that simulations of the full photothermal coupling would require complex coupled structural, thermal and optical FEM simulations and would necessitate a priori or independently experimentally determined knowledge of the photoelastic coefficient and thermalization time for the specific material and structure investigated. The simulations in this work thus only test the predictions of the purely dispersive Fano optomechanics model ($\beta=0$).

For the broadband mirror cavity consisting of the Bragg mirror and the high-refractive index grating the total round-trip loss is 8\%. Since the previously defined infinite subwavelength grating mirror would display unity peak reflectivity at $\lambda_0$ under plane wave illumination, we add absorption losses by attributing a small imaginary part ($4.8\times 10^{-4}$) to the grating refractive index, so that the total round-trip loss of the Fano cavity is the same as the broadband cavity.

For each cavity length such that the cavity resonates at $\lambda_0$ the cavity spectrum is simulated and the linewidth extracted. The Fano mirror is then translated by 10 pm and the cavity resonance shift is calculated by evaluating the displacement sensitivity 
\begin{equation}
S_\lambda=\frac{d\lambda_0}{dq}=-\frac{T_{\lambda q}}{T_{\lambda \lambda}},
\end{equation} 
where $T_{\lambda q}$ is the mixed derivative, evaluated using a central multivariate finite-difference scheme, and $T_{\lambda \lambda}$ is the second derivative with respect to the wavelength.

\begin{figure}[h]
\includegraphics[width=0.50\columnwidth]{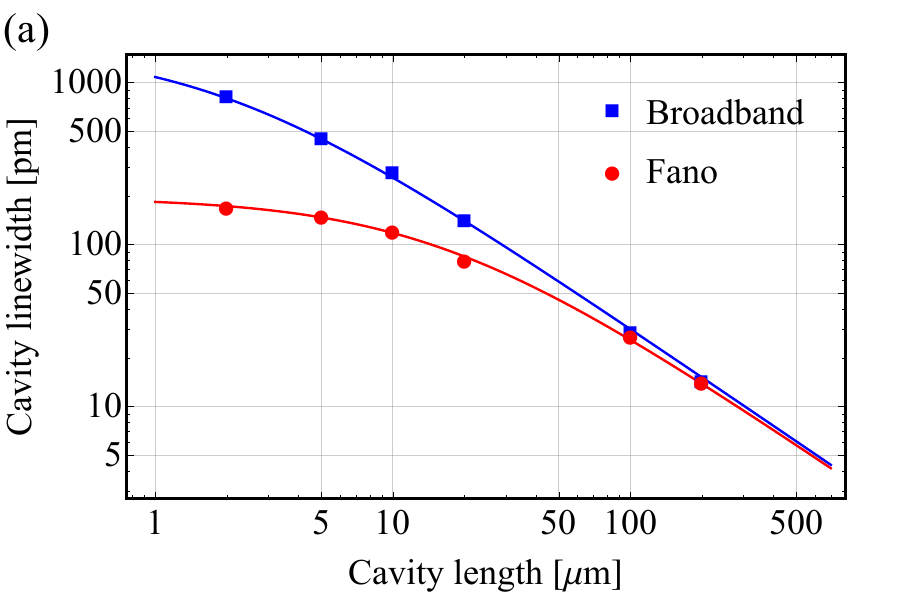}
\includegraphics[width=0.49\columnwidth]{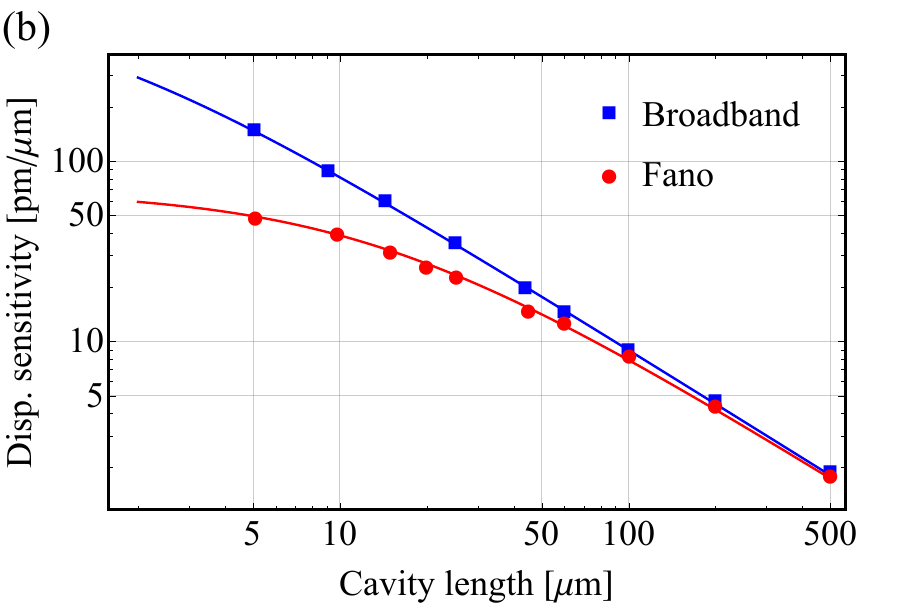}
\caption{Simulated cavity linewidth (a) and displacement sensitivity $S_\lambda$ (b) as a function of cavity length $L$. The blue squares and the red dots show the results for the broadband mirror and Fano mirror cavities, respectively. The solid lines show results of fits of the form $A/(L+L_0)$.}
\label{fig:PBC_results}
\end{figure}

The results of the simulated cavity linewidths and displacement sensitivities as a function of cavity length are shown in Fig.~\ref{fig:PBC_results}. The expected saturation of the linewidth and the displacement sensitivity at short lengths is clearly visible for the Fano cavity, in contrast with the broadband cavity, and observed to be in agreement with the results of fits of the form $A/(L+L_0)$. The values of the extracted effective lengths are reported in Table~\ref{tab:L0} and compared to the predictions of the analytical model of Ref.~\cite{Mitra2024}
\begin{equation}
\label{eq:L0thFano}
L_0^\textrm{th}=\frac{2(1-r_d)\lambda_0^2}{8\pi\gamma_\lambda},
\end{equation}
where the resonant wavelength $\lambda_0=938.7$ nm, the linewidth (HWHM) in wavelength units of the Fano mirror $\gamma_\lambda=3$ nm and its off-resonant reflectivity level (in amplitude)  $r_d=0.59$ are obtained from a fit to the simulated Fano mirror transmission spectrum.

For the broadband cavity the dispersion of the mirrors can be taken into account to evaluate the effective cavity length correction due to the penetration of the field into the mirrors and generally can be written~\cite{Garmire2003}
\begin{equation}
\label{eq:L0thB}
L_0^\textrm{th}=\frac{\lambda_0^2}{4\pi}\frac{d\phi}{d\lambda},
\end{equation}
where $\phi$ is the phase of the reflectivity coefficient. For a Bragg mirror such as the incoupling mirror considered here, the dispersion can be calculated analytically from the refractive index and thicknesses of the alternating dielectric layers~\cite{Garmire2003}, yielding here $L_0^\textrm{th}\simeq 1.3$ $\mu$m.

\begin{table}
\caption{Values of the parameter $L_0$ extracted from fits with $A/(L+L_0)$. The results are denoted by $L_0^{\delta\lambda}$, obtained by fitting the cavity linewidth data in Fig.~\ref{fig:PBC_results}(a), and $L_0^{S_\lambda}$, obtained by fitting the displacement sensitivity data in Fig.~\ref{fig:PBC_results}(b). The values of the theoretical estimates for both broadband mirror and Fano mirror cavities with the same total losses obtained from Eqs.~(\ref{eq:L0thFano}) and (\ref{eq:L0thB}) are denoted by $L_0^\textrm{th}$.}
  \label{tab:L0}
  \begin{tabular}{|c|ccc|}
   \hline
   Cavity & $L_0^{\delta\lambda}$ ($\mu$m) & $L_0^{S_\lambda}$ ($\mu$m) & $L_0^\textrm{th}$  ($\mu$m)\\\hline
   Broadband & 1.8 & 1.8 & 1.3 (Bragg)\\
   Fano & 12 & 11 & 9.5 \\\hline
  \end{tabular}
\end{table}

The simulations performed for the suspended subwavelength grating mirror thus confirm the main conclusion of the analytical model, which is that the reduction in a Fano cavity linewidth in the Fano regime is conccomitant with a similar reduction in optomechanical coupling. This conclusion also holds when finite grating/beam size effects and nonplanar deformations of the mechanical structure are considered, as shown in the Appendix.

%%%%%%%%%%%%%%%%%%%%%%%%%%%%%%%%%%%%%%%%%%%%%%%%%%%%

\section{Experimental methods and results}
\label{sec:experiment}

\subsection{Suspended Fano mirror}

\begin{figure}[h]
\begin{subfigure}{0.49\textwidth}
\centering\includegraphics[width=\columnwidth]{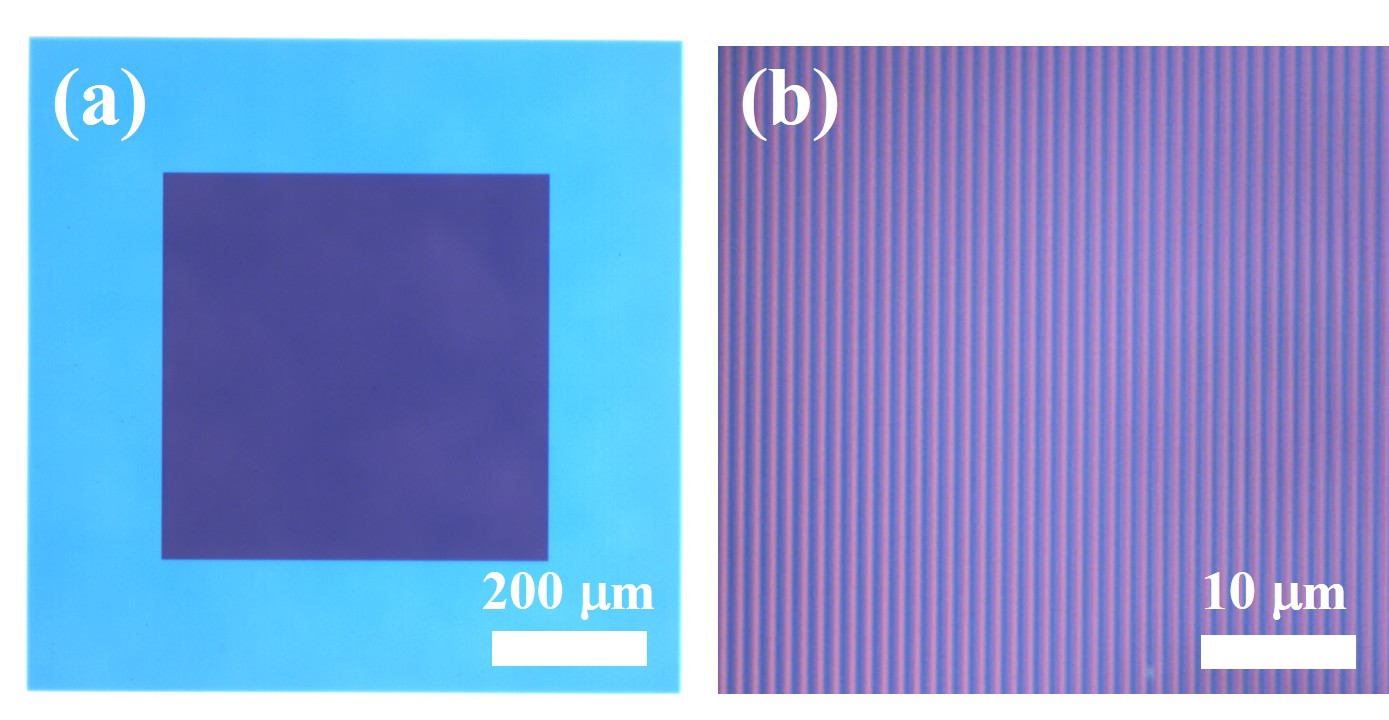}
\begin{minipage}{.3cm}
\vfill
\end{minipage}
\end{subfigure}
\begin{subfigure}{0.49\textwidth}
\centering\includegraphics[width=\columnwidth]{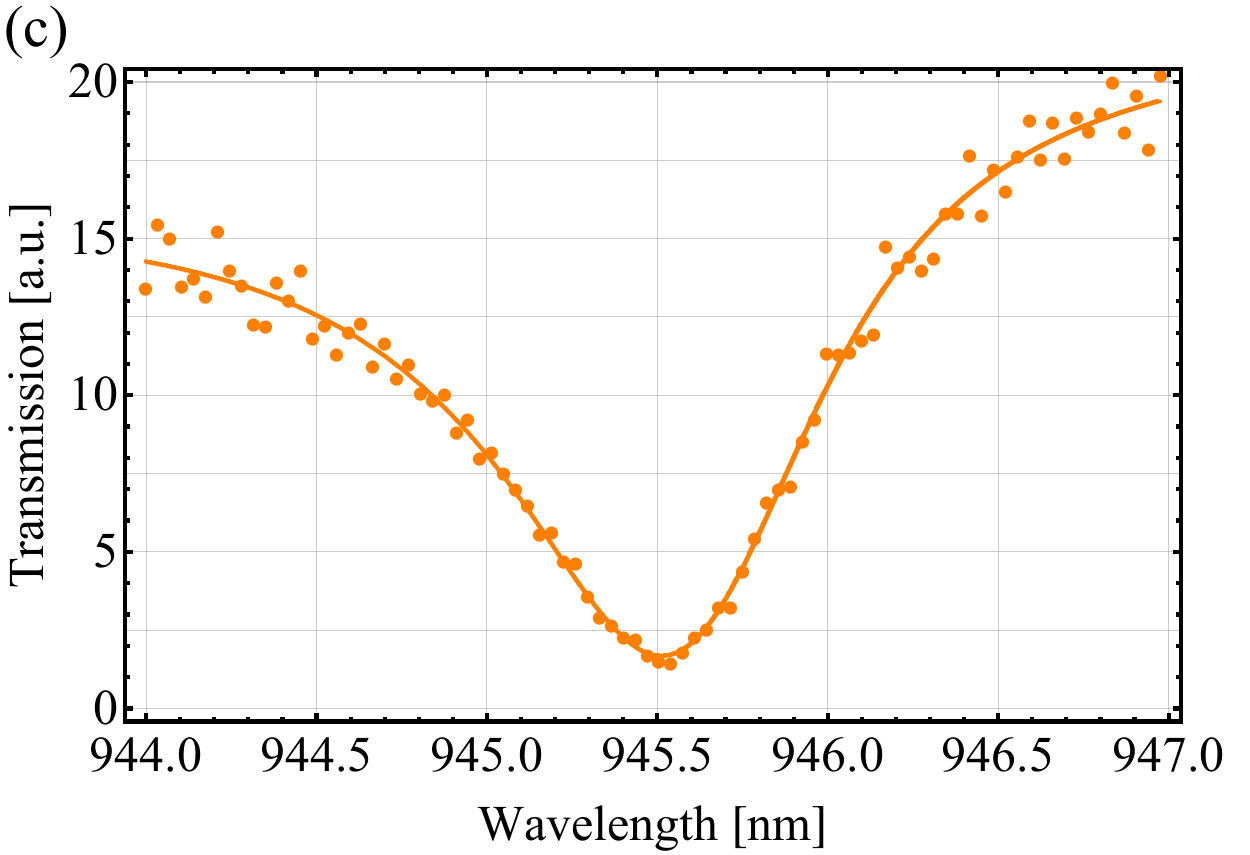}
\end{subfigure}
\caption{(a) Top-view picture of the suspended Fano mirror. (b) Zoom-in picture of the patterned area. (c) Transmission spectrum of the Fano mirror used for the experimental investigations. The solid line shows the result of a fit with the theoretical model.}
\label{fig:grating_exp}
\end{figure}

The Fano mirror used for the experimental investigations is similar to the "high-Q" grating used in Ref.~\cite{Mitra2024}. It consists in a 1 mm-square, 153 nm-thick SiN membrane suspended on a 5 mm-square, 500 $\mu$m-thick Si frame and patterned with a 600 $\mu$m-square subwavelength grating. Figure~\ref{fig:grating_exp} shows a transmission spectrum of the grating illuminated at normal incidence by a TM-polarized, 260 $\mu$m-waist Gaussian beam. The off-resonant reflectivity $|r_d|^2$ is $0.35$. The reflectivity resonance HWHM is $0.5$ nm and the peak reflectivity at 945.5 nm is 92\% ($T_0\simeq 4\%$, $A\simeq 4\%$). While the intrinsic absorption of the low-stress SiN films used here is known to be very low~\cite{Land2024}, extra losses arise from the imperfect interference between the incident mode and the guided mode in the grating~\cite{ToftVandborg2021,Mitra2024}, which limit the peak reflectivity of the Fano mirror. With these parameters the effective length $L_0$ given by Eq.~(\ref{eq:L0thFano}) is then about 58 $\mu$m.

\subsection{Fano cavity setup}

\begin{figure}[h]
\centering
\includegraphics[width=0.7\columnwidth]{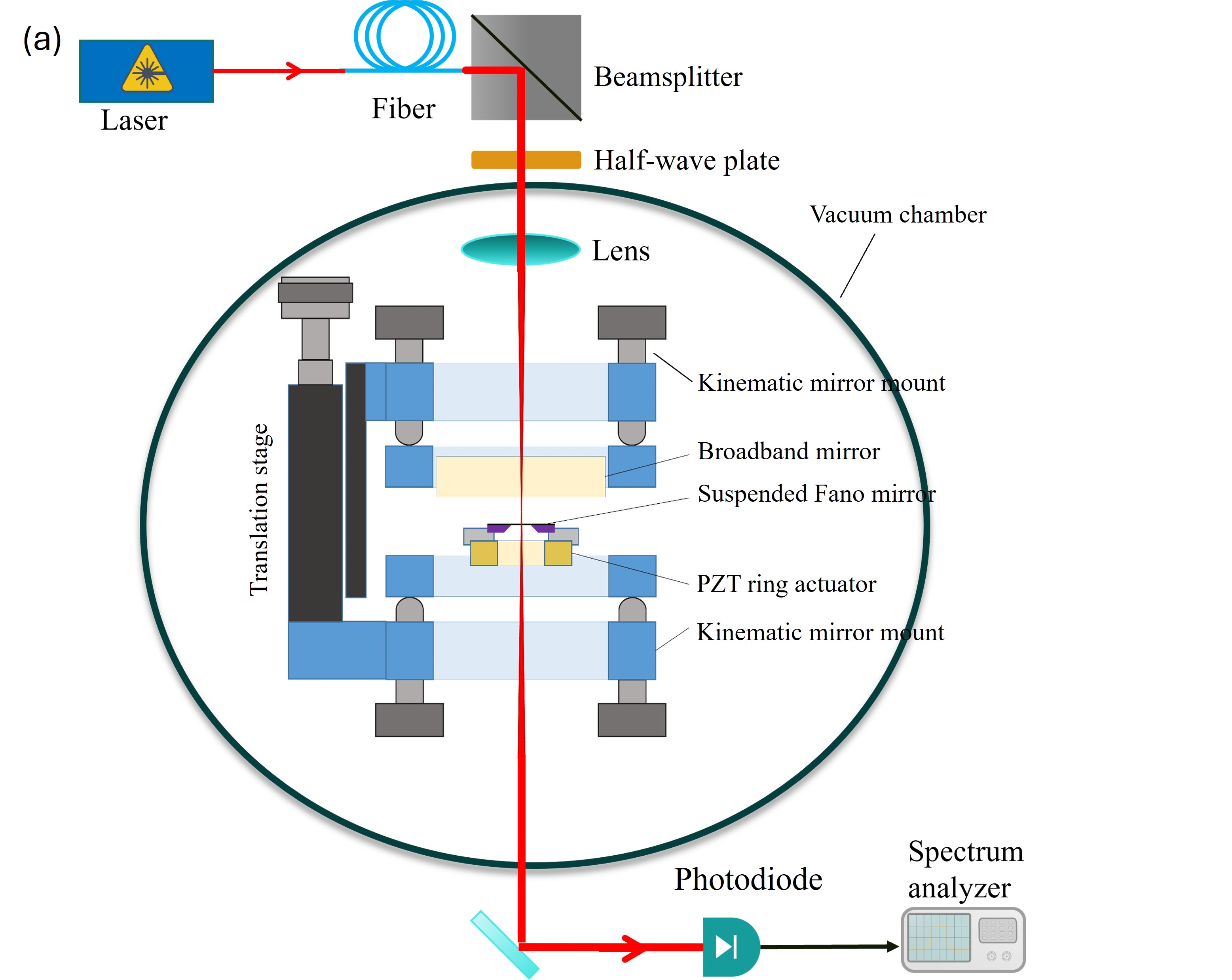}\\\vspace{0.3cm}
\includegraphics[width=0.49\columnwidth]{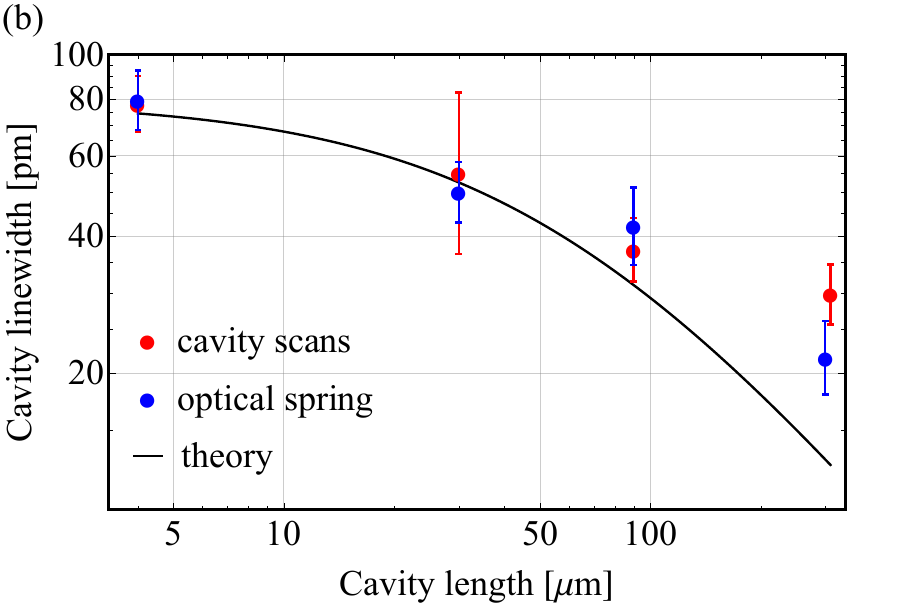}\includegraphics[width=0.49\textwidth]{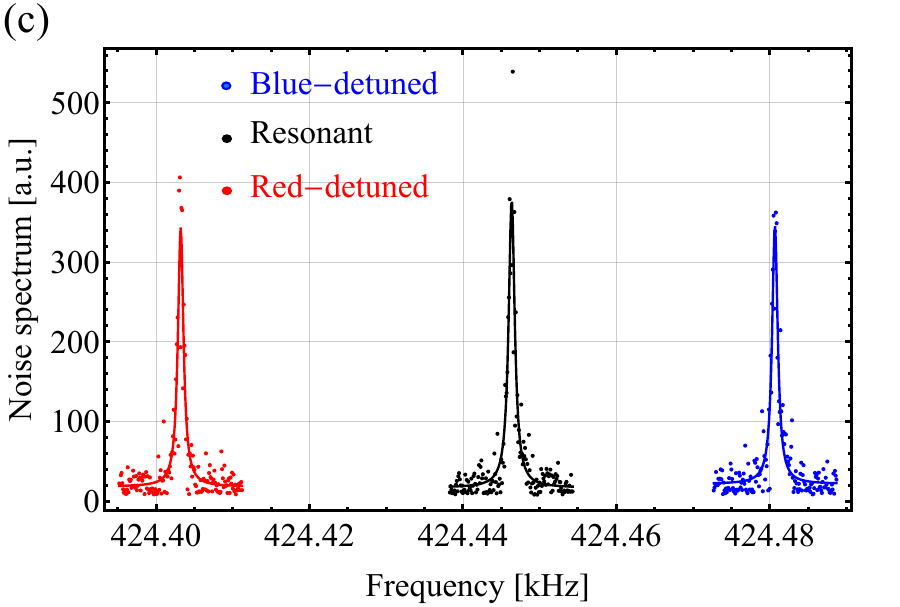}
\caption{(a) Skematic of the experimental setup. (b) Cavity linewidths (HWHM) as a function of cavity length. The red and blue dots respectively show the linewidths extracted from the cavity transmission spectra and from fits to the optical spring measurements shown in Fig.~\ref{fig:results_detuning}. The solid line shows the theoretical predictions. (c) Example of thermal noise spectra used for the measurement of the optical spring frequency shift. For this measurement, the cavity length was 300 $\mu$m, the input power 100 $\mu$W and the detunings (in wavelength units) from resonance (black) were $+70$ pm (red) and $-59$ pm (blue).}
\label{fig:setup}
\end{figure}

To investigate Fano cavity optomechanics with this Fano mirror we make use of a setup (Fig.~\ref{fig:setup}(a)) very similar to that used in Ref.~\cite{Mitra2024}, the main difference being that the Fano cavity is placed inside a vacuum chamber ($\sim10^{-6}$ mbar), so that the effect of air damping on the vibrations of the Fano mirror can be ignored. In brief, monochromatic light from a tunable laser is injected into the cavity consisting of a broadband mirror (reflectivity 96\% at 945.5 nm) and the Fano mirror. The finesse of the corresponding broadband cavity is thus about 40. The parallelism and distance between the two mirrors can be precisely controlled using piezo-actuated mirror mounts and translation stages. The light transmitted is collected by a low-noise photodetector and the signal is processed with a high-resolution spectrum analyzer. 

We investigate the optical spring effects for 4 different lengths of the Fano cavity, ranging from 4 to 300 $\mu$m. For each experiment the cavity length is adjusted so that the cavity resonates at the Fano mirror resonance wavelength. The cavity length is then calculated via a measurement of the cavity free-spectral range from a broadband wavelength scan. The cavity linewidth can be determined by short wavelength scans around resonance and fits with the theoretically expected Fano profiles (Eq.~(\ref{eq:trans_ana})). The variation of the linewidths with the cavity length is shown in Fig.~\ref{fig:setup}(b), together with the theoretical predictions based on the experimentally determined grating transmission model parameters and cavity lengths. Figure~\ref{fig:setup}(b) also shows the cavity linewidths determined from optical spring measurements which will be discussed in Sec.~\ref{sec:experimental_results}. While a good agreement between the experimentally determined linewidths and the theoretical predictions is observed at short lengths, somehow broadened linewidths are observed at longer cavity lengths due to the higher sensitivity of the cavity to imperfect alignment and vibrations, as discussed in~\cite{Mitra2024,Kirkegaard2025}.

Once the cavity length is adjusted and the cavity resonance wavelength and linewidth are determined, thermal noise spectra are recorded for various powers and/or wavelengths of the light injected into the cavity, as discussed in detail in the next section. We focus primarily on measurements of the mechanical resonance shift of the (2,2) mode at 420 kHz of the Fano mirror (mechanical Q-factor of $3\times 10^6$ in vacuum). An example of such a measurement is shown in Fig.~\ref{fig:setup}(c) taken for a cavity length of 300 $\mu$m, an input power of 100 $\mu$W and different cavity detunings. The optical spring shift can then be determined as the difference in resonance frequencies obtained from Lorentzian fits of these spectra. No noticeable change in the mechanical linewidth was observed, as expected from our operating deeply in the unresolved sideband regime.

\subsection{Experimental results}
\label{sec:experimental_results}

\begin{figure}[h]
\centering
\includegraphics[width=\columnwidth]{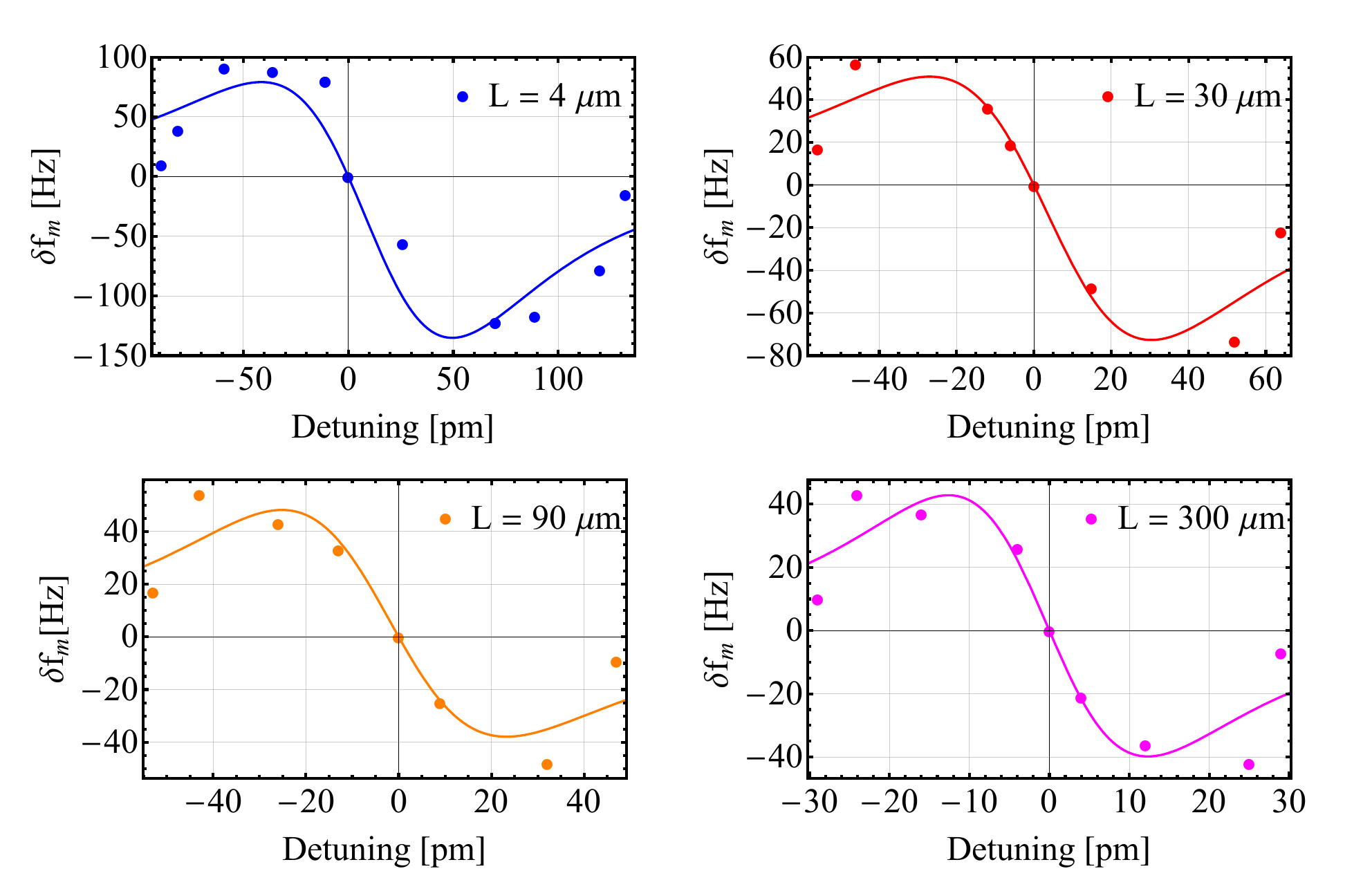}
\caption{Experimental variations of the optical spring frequency shift as a function of detuning (in wavelength units) for a fixed input power of 100 $\mu$W and for the 4 cavity lengths. The lines represent the results of fits with Eq.~(\ref{eq:deltafm_detuning}).}
\label{fig:results_detuning}
\end{figure}

Two types of measurements were then performed: first, the input power to the cavity was fixed (100 $\mu$W) and the mechanical resonance shift of the (2,2) mode of the Fano mirror is determined for various detunings. The results of the optical spring shift as a function of the cavity detuning for the 4 different cavity lengths are shown in Fig.~\ref{fig:results_detuning}, together with the results of fits of the form (Eq.~(\ref{eq:shift_a_ana}))
\begin{equation}
\label{eq:deltafm_detuning}
\delta f_m=\frac{\delta\omega_m}{2\pi}=\frac{(\alpha \eta)^2}{2\pi} \frac{P_\textrm{in}}{\hbar\omega_0}\frac{c^3T_R}{(L+L_0)^3}\frac{\Delta}{(\Delta^2+\kappa_F'^2)^2},
\end{equation}
where we have introduced the factor $\alpha=G_\textrm{exp}/G$ is a free parameter relating the measured optomechanical coupling $G_\textrm{exp}$ to the theoretically expected dispersive linear optomechanical coupling $G=\eta\kappa_L$, $P_\textrm{in}$ is the input power to the cavity. The expected dispersive linear coupling can be evaluated using an effective mass $m_\textrm{eff}=m/4$, where $m=\rho t a$ is the physical mass of the membrane, $\rho=3100$ kg/m$^3$ being its density, which was determined in~\cite{Darki2022}, $t=153$ nm its thickness and $a=1$ mm its transverse size. This yields $m_\textrm{eff}\simeq 123$ ng, a zero-point motion amplitude of $q_0=0.4$ fm and a Lamb-Dicke parameter of $\eta\simeq 2.7\times 10^{-9}$ for the (2,2)-mode at $f_m=420$ kHz. 

For the fits the cavity linewidth $\kappa_F$ was left as a free-parameter, and the resulting values, shown in Fig.~\ref{fig:setup}(b), are observed to be consistent with the values previously determined during the alignment of the cavity at the chosen length. The values of $\alpha$ resulting from the fits are reported in Table~\ref{tab:alphas} and are observed to be approximately independent of the cavity length, as expected, but considerably much larger than unity, which indicates a much stronger optomechanical interaction than expected from the  dispersive only optomechanics model. Indeed, for a cavity length of 4 $\mu$m one gets a dispersive linear optomechanical coupling $G=(2\pi)$ 31.7 kHz, which, for an input power of 100 $\mu$W, would give a maximum optical spring shift of 0.3 Hz, which is about 400 times less than what is experimentally observed, and which is consistent with a value of $\alpha\simeq 20$, since $\alpha^2$ corresponds to the factor $\xi=1+\beta$ in the theoretical model.

\begin{table}
\caption{Fit results for $\alpha$ from the detuning and power measurements for the 4 cavity lengths.}
  \label{tab:alphas}
  \begin{tabular}{|ccc|}
   \hline
   Cavity length $L$ ($\mu$m) & $\alpha$ (detuning fit) &   $\alpha$ (power fit) \\\hline
    4 & $24.1\pm 4.9$ & $21.8\pm 3.2$\\
	30 & $15.5\pm 3.3$ & $16.3\pm 2.1$\\
	90 & $21.5\pm 5.8$ & $25.4\pm 3.9$\\
    300 & $29.8\pm 7.3$ & $37.2\pm 5.5$ \\\hline
  \end{tabular}
\end{table}

%%%%%%%%%%%%%%%%%%%%%%%%%%%%%%%%%%%%%%%%%%%%%%%%%%

\begin{figure}[h]
\centering
\includegraphics[width=0.49\columnwidth]{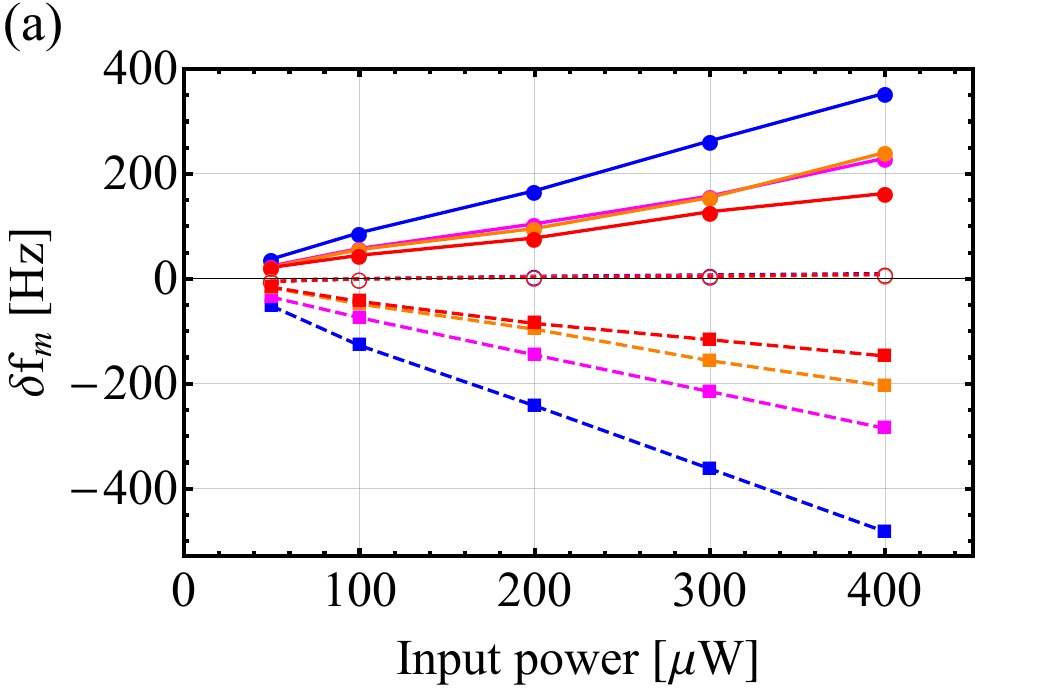}
\includegraphics[width=0.49\columnwidth]{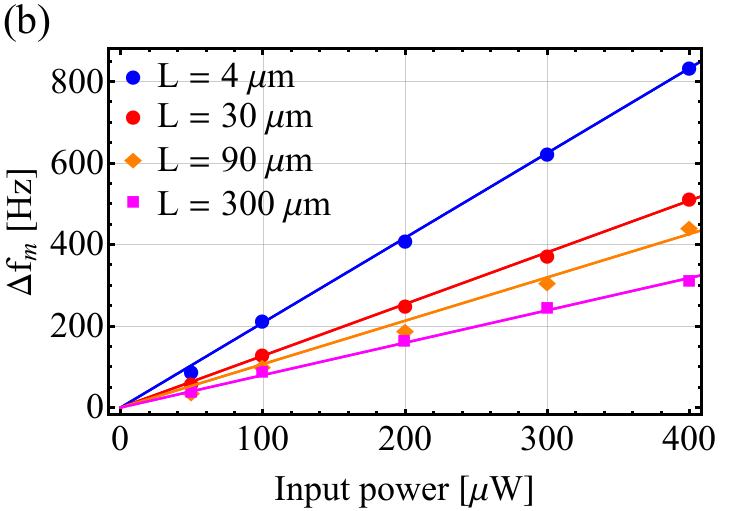}
\caption{(a) Experimental variations of the optical spring shift with input power for the 4 cavity lengths and for red (full circles), zero (empty circles) and blue (squares) detunings. The blue and red detunings (in wavelength units) are $-59,-46,-43,-24$ pm and $70,52,43,32$ pm (red), respectively, for $L=(4,30,90,300)$ $\mu$m. The lines are guides for the eye. (b) Variations of the difference of the red- and blue-detuned shifts $\Delta f_m$ as a function of input power for the 4 cavity lengths. The lines are the results for each cavity lengths of fits with Eq.~(\ref{eq:Deltafm}).}
\label{fig:results_power}
\end{figure}

Second, the input laser was detuned by approximately a half-linewidth to the red or blue of the cavity resonance and the mechanical resonance frequency shift was measured for different input powers. As a reference, the mechanical resonance frequency shift when the laser was tuned to resonance with the cavity was also measured. The results of the detuned measurements, shown in Fig.~\ref{fig:results_power}(a), display the expected linear variations with the power, whereas the resonance measurements show almost no shift in comparison, indicating that thermal shifts due to static absorption are negligible. To evaluate the strength of the optomechanical coupling (i.e. $\alpha$) we performed fits of the difference of the shifts measured with the blue and red detunings, $\Delta_b$ and $\Delta_r$, for the same power (shown in Fig.~\ref{fig:results_power}(b)) anf for a given cavity length
\begin{equation}
\Delta f_m(P_\textrm{in})=\delta f_m(\Delta_b)-\delta f_m(\Delta_r)=\frac{(\alpha\eta)^2}{2\pi}\frac{P_\textrm{in}}{\hbar\omega_0}\frac{c^3T_R}{(L+L_0)^3}\left[\frac{\Delta_r}{(\kappa_F'^2+\Delta_r^2)^2}-\frac{\Delta_b}{(\kappa_F'^2+\Delta_b^2)^2}\right].
\label{eq:Deltafm}
\end{equation}
The values of $\alpha$ resulting from the fits using for the linewidths the best fit results from the detuning measurements are reported in Table~\ref{tab:alphas} for the 4 cavity lengths. They are observed to consistent with those obtained from the detuning measurements, and clearly indicate a much stronger optomechanical coupling than the expected dispersive linear coupling.

\begin{figure}[h]
\centering
\includegraphics[width=0.6\columnwidth]{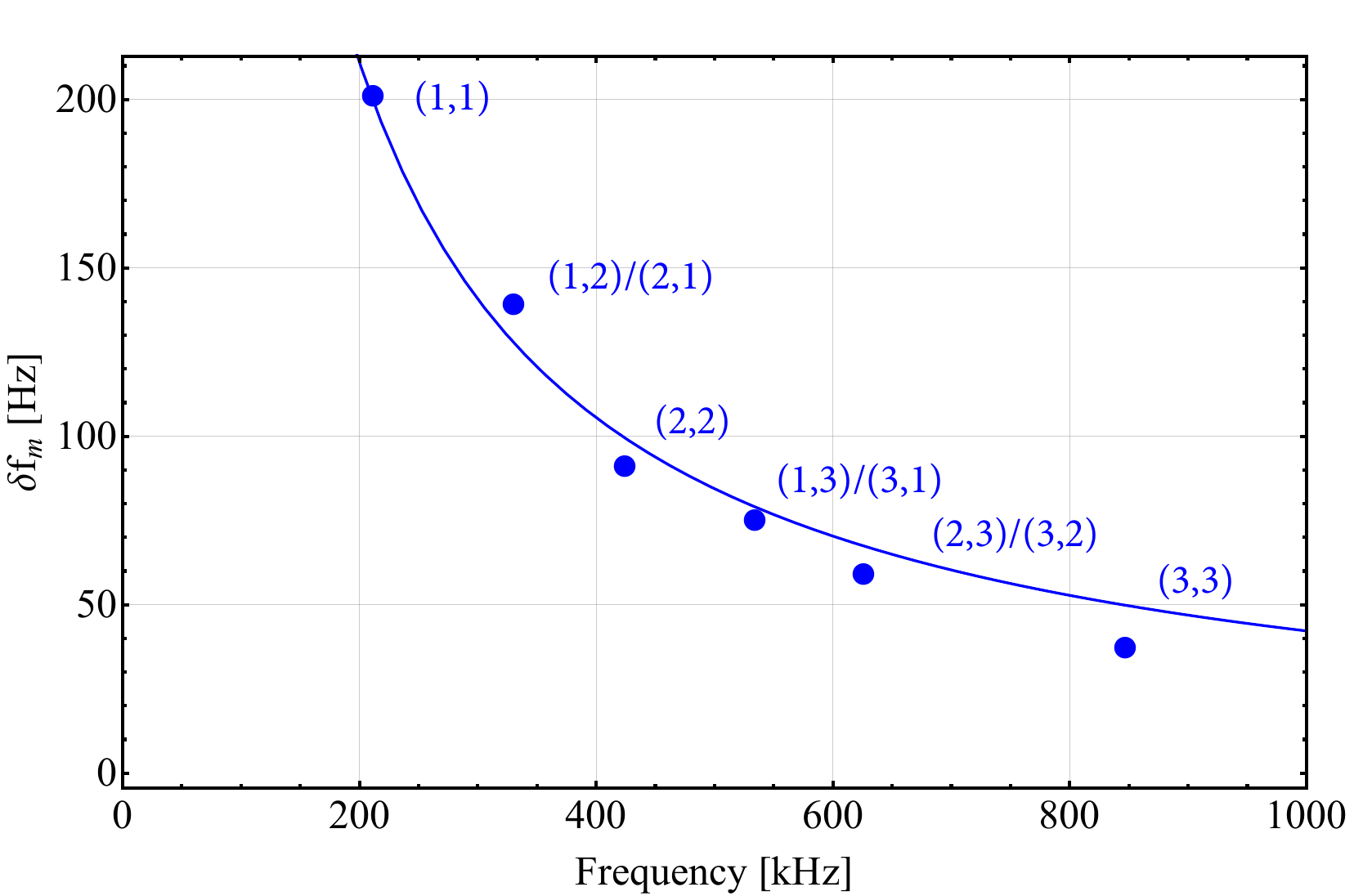}
\caption{Variation of the optical spring of different mechanical modes for a 4 $\mu$m-long cavity, an input power of 100 $\mu$W and a blue detuning of $-60$ pm. The $(m,n)$ indices refer to the corresponding square drum mode indices. The line shows the result of a fit with $1/f_m$.}
\label{fig:modes}
\end{figure}

For completness, Figure~\ref{fig:modes} shows the variation of the optical spring for the lowest vibrational modes of the Fano mirror for a 4 $\mu$m-long cavity, an input power of 100 $\mu$W and a blue detuning of $-60$ pm. The magnitude of the optical spring is approximately proportional with $1/f_m$, as expected from the scaling of the Lamb-Dicke parameter with the mechanical frequency.

%%%%%%%%%%%%%%%%%%%%%%%%%%%%%%%%%%%%%%%%%%%%%%%%%%%%

\section{Conclusion}
\label{sec:conclusion}
In this work we investigated cavity optomechanics in Fano cavities, i.e. cavities possessing a mirror with a high-Q internal optical resonance. First, we introduced a generic Fano cavity optomechanics model based on the generalized input-output theory of Ref.~\cite{Cernotik2019} for cavities with strongly wavelength-dependent reflectors and which includes both dispersive and photothermal optomechanical couplings. Analytical expressions for the modifications of the mechanical frequency and damping of the vibrating Fano mirror due to the radiation pressure force were derived with a focus on the unresolved sideband regime relevant to this work. 

An important conclusion for this type of Fano cavities is that the linewidth narrowing of such a Fano cavity at short lengths as compared to the corresponding broadband mirror cavity is accompanied by a concomitant reduction in the dispersive optomechanical coupling, such that, for instance, the variations of the optical spring shift with input power or cavity detuning are comparable to those of a broadband mirror cavity. These generic results were corroborated by FEM simulations of realistic Fano mirror structures based on thin, suspended subwavelength gratings.

We then investigated experimentally the optomechanical response of Fano cavities consisting of a plane-plane arrangement of a standard broadband reflectivity mirror and a suspended SiN subwavelength grating membrane for cavity lengths in the range of a few to a few hundreds of microns. The measured optical spring variations with the input light power and detuning were observed to match those expected from the Fano cavity optomechanics model including dispersive and photothermal effects, the magnitude of the latter being several hundred times larger than the former.

While these results call for further investigations of photothermal effects in such photonic crystal membranes, the observed strong optomechanical interactions combined with new possibilities for optomechanical control~\cite{Cernotik2019,Fitzgerald2021,Peralle2024,Du2026} already suggest that Fano cavities with suspended resonant mirrors are promising for quantum optomechanics with membrane-at-the-end cavities~\cite{Sang2022,Xu2022,Zhou2023,Enzian2023,Khokhar2026}.

%%%%%%%%%%%%%%%%%%%%%%%%%%%%%%%%%%%%%%%%%%%%%%%%%%%%%

\section*{Appendix}

The infinite grating/plane wave model of Sec.~\ref{sec:PBCmodel} does not take into account the collimation and finite size of the beam impinging on a finite-size grating, which are known to potentially affect the optical properties of Fano mirrors~\cite{ToftVandborg2021} and Fano cavities~\cite{Naesby2018,Mitra2024}. Moreover, by construction, the model only allows one to evaluate a displacement sensitivity corresponding to a translation of the Fano mirror, whereas in the experiments reported here with suspended Fano mirror the displacement is due to a deformation of the structure and the optomechanical coupling depends on the overlap between the optical mode and the mechanical mode considered~\cite{Pinard1999}. 

\begin{figure}[h]
\includegraphics[width=0.49\columnwidth]{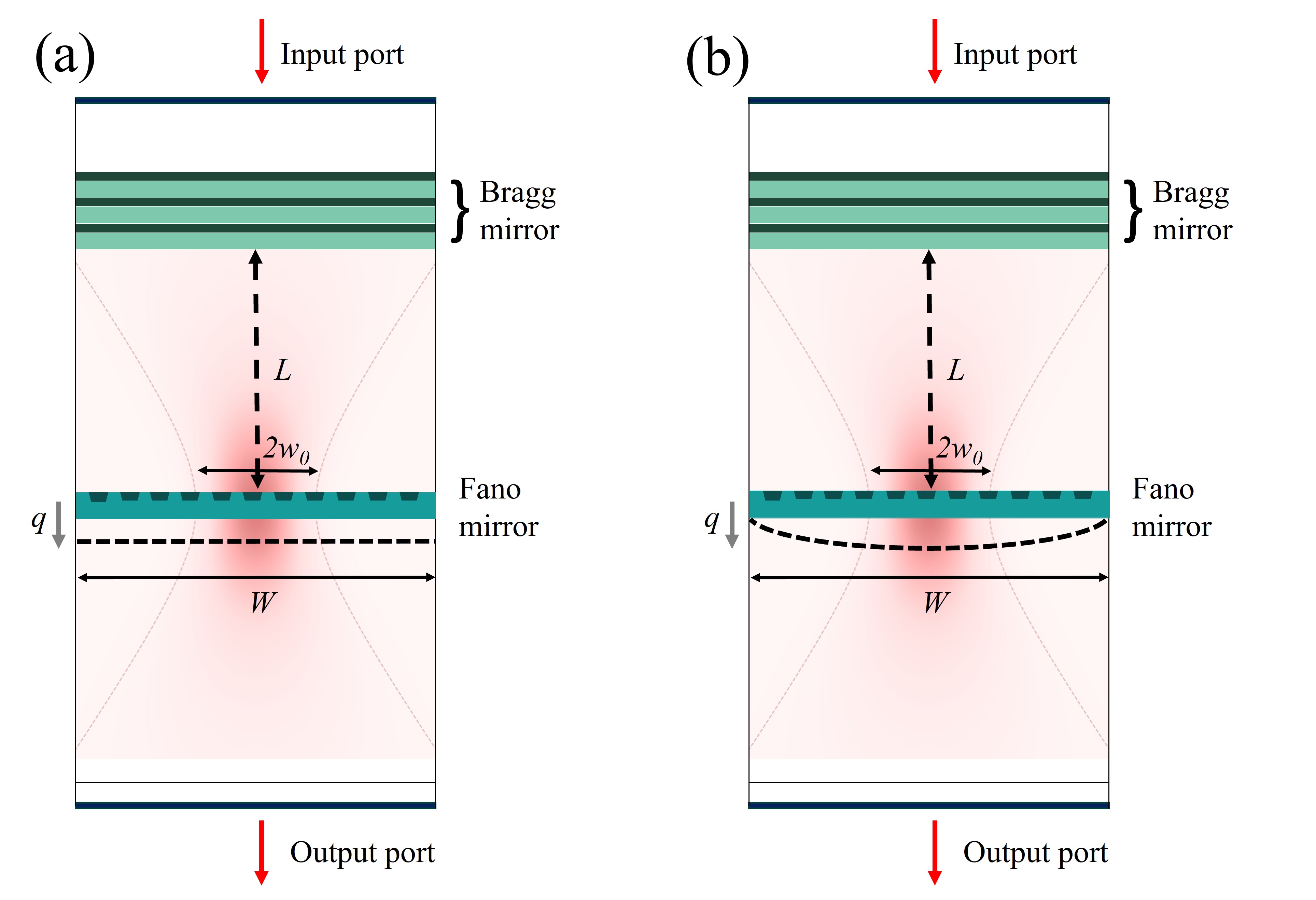}
\includegraphics[width=0.50\columnwidth]{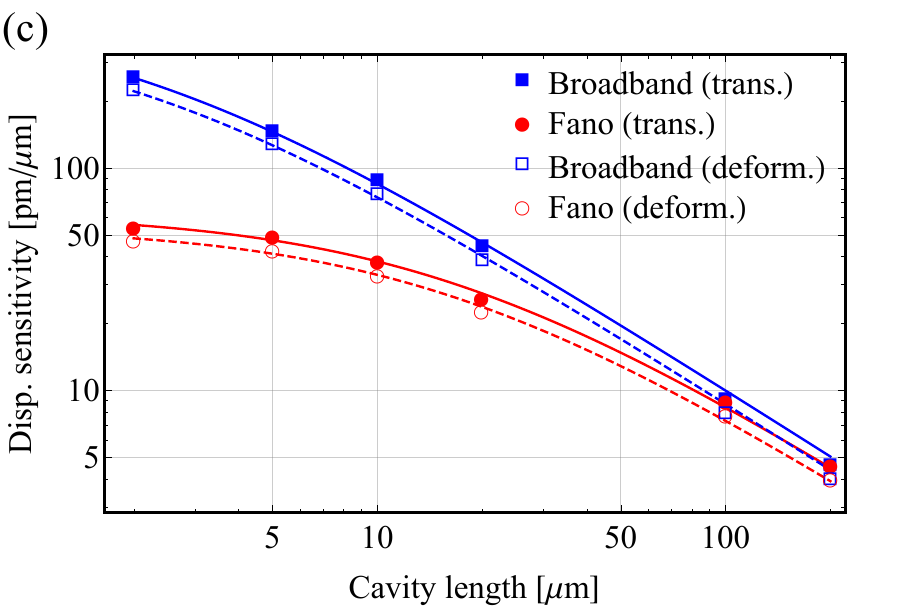}
\caption{Gaussian beam/finite size mirror models for optomechanical cavities in which the Fano mirror is translated (a) and deformed (b). (c) Simulated displacement sensitivities as a function of cavity length for broadband mirror (blue squares) and Fano mirror (red dots) cavities. The full and empty symbols refer respectively to the cases where the Fano mirror is translated and deformed. The solid and dashed lines show the results of fits of the form $A/(L+L_0)$.}
\label{fig:fullmodel}
\end{figure}

FEM simulations were thus performed with a finite-size grating with width $W$ illuminated by a Gaussian beam with waist $w_0$ focused on the grating. Perfectly matched layers were used as boundaries. The beam waist in the simulations was chosen to be 40 $\mu$m, so that the peak reflectivity of the (non-absorbing) grating is approximately 96\%, due to finite-size and collimation effects. The size of the grating is taken to be $W=3w_0=120$ $\mu$m, so that diffraction and finite-grating size losses are negligible. Two sets of simulations of the displacement sensitivity were performed: by vertically translating the Fano mirror by a fixed amount $\bar{q}$ (e.g. 10 pm, Fig.~\ref{fig:fullmodel}(a)) and by deforming the grating vertically with a deformation $\bar{q}\cos(\pi x/W)$ corresponding to the fundamental out-of-plane drum mode (Fig.~\ref{fig:fullmodel}(b)). In this case, one expects that the displacement sensitivity is reduced by a factor given by the overlap between the mechanical and optical modes
\begin{equation}
\int_{-W/2}^{W/2}dx\,\cos\left(\frac{\pi x}{W}\right)\sqrt{\frac{2}{\pi w_0}}\exp\left(-\frac{2x^2}{w_0^2}\right),
\end{equation}
which, for the parameters considered, is about 87\%. This is indeed what is observed in Fig.~\ref{fig:fullmodel}(c), which shows the displacement sensitivity for both cavities and both types of displacements. These simulations also confirm that there is no additional dispersive optomechanical coupling with the guided mode when the structure is deformed according to one of its normal out-of-plane modes, and thus supports the applicability of the coupled-mode model of Sec.~\ref{sec:theory} to realistic structures, such as the suspended subwavelength grating mirror investigated experimentally.

%\begin{backmatter}
%\bmsection{Funding}
\section*{Funding}
Novo Nordisk Fonden.

%\bmsection{Acknowledgments}
\section*{Acknowledgments}
We are grateful to Claudiu Genes for enlightening discussions regarding the Fano cavity optomechanics model and Ali A. Darki and John E. V. Andersen for their contributions to the design of the vacuum cavity system.

%\bmsection{Disclosures}
\section*{Disclosures}
The authors declare no conflicts of interest.

%\bmsection{Data Availability Statement}
\section*{Data Availability Statement}
Data underlying the results presented in this paper are not publicly available at this time but may be obtained from the authors upon reasonable request.

%\bmsection{Supplemental document}
%\section*{Supplemental document}
%See Supplement 1 for supporting content. 

%\end{backmatter}

\end{document}